\documentclass{article} 
\usepackage{GEM_workshop_2025,times}

\usepackage{amsmath,amsfonts,bm}

\def\eqref#1{equation~\ref{#1}}

\def\1{\bm{1}}

\DeclareMathAlphabet{\mathsfit}{\encodingdefault}{\sfdefault}{m}{sl}
\SetMathAlphabet{\mathsfit}{bold}{\encodingdefault}{\sfdefault}{bx}{n}

\usepackage{hyperref}
\usepackage{url}
\usepackage{graphicx}
\usepackage{subcaption}
\usepackage{float}
\usepackage{booktabs}
\usepackage{dsfont}
\title{Sesame: Opening the door to protein pockets}

\author{
Raúl Miñán$^{1*}$, Carles Perez-Lopez$^{2}$, Javier Iglesias-Fernandez$^{3}$, Álvaro Ciudad$^{1*}$, Alexis Molina$^{1*}$ \\
\\
$^{1}$Department of Artificial Intelligence, Nostrum Biodiscovery, Barcelona, Spain \\
$^{2}$Department of Drug Discovery, Nostrum Biodiscovery, Barcelona, Spain \\
$^{3}$Department of R\&D, Nostrum Biodiscovery, Barcelona, Spain \\    
 *\texttt{\{raul.minan,alvaro.ciudad,alexis.molina\}@nostrumbiodiscovery.com} \\
}

\iclrfinalcopy 
\begin{document}

\maketitle

\begin{abstract}

Molecular docking is a cornerstone of drug discovery, relying on high-resolution ligand-bound structures to achieve accurate predictions. However, obtaining these structures is often costly and time-intensive, limiting their availability. In contrast, ligand-free structures are more accessible but suffer from reduced docking performance due to pocket geometries being less suited for ligand accommodation in apo structures. Traditional methods for artificially inducing these conformations, such as molecular dynamics simulations, are computationally expensive. In this work, we introduce Sesame, a generative model designed to predict this conformational change efficiently. By generating geometries better suited for ligand accommodation  at a fraction of the computational cost, Sesame aims to provide a scalable solution for improving virtual screening workflows.

\end{abstract}

\section{Introduction}
Within the realm of drug discovery, molecular docking serves as a pivotal methodology for exploring and characterizing the intricate interactions between molecular entities and their corresponding protein binding cavities \citep{Meng2011-at}. Traditionally, high-resolution ligand-bound, or holo structures, are needed as the starting point for this computational workflow \citep{docking_guidelines}. However, they might not be always available, as they are produced by costly experimental protocols \citep{x_ray_mh}. 

Ligand-free, or apo structures, are more readily available. They can be derived from experimental methods, similar to those used for holo structures, or predicted employing models like AlphaFold2 \citep{alphafold2}. Despite their availability, when used as input for virtual screening campaigns, they suffer from lower performance compared to holo structures, both in traditional docking \citep{vs_alphafold} and deep learning approaches \citep{diffdock}. This performance drop is attributed to pocket geometries being better suited for ligand accommodation in holo structures. This problem is particularly relevant when transient binding sites, such as allosteric and cryptic pockets, are of interest. These unique pockets are binding sites that only become apparent under an induced conformational change, e.g. drug binding, providing a promising alternative to classical binding sites for drug development \citep{another_enhanced}. Moreover, they can be involved in mechanisms that allow for additional specificity in modulating certain diseases \citep{cryptic_allostery}.

Traditional approaches for generating these holo-like geometries consist in computationally expensive molecular dynamics (MD) protocols that can take up to several GPU-days \citep{mixed_solvent}. In this work we present Sesame, a generative model that can generate these desired geometries for a fraction of the cost by predicting the apo-holo conformational changes in a ligand-agnostic way.

\section{Related Work}
\textbf{Traditional Molecular Dynamics Approaches}. MD simulations are computational methods that calculate atomic and molecular interactions and physical movements over time by solving Newton's equations of motion. These simulations can be used to characterize the formation of cryptic pockets through numerous approaches: standard MDs with organic probe molecules to measure binding propensity of fragments \citep{organic_probe}, mixed-solvent MDs to characterize hydrophobic regions \citep{mixed_solvent} and enhanced sampling methods to accelerate the sampling of the conformations of interest \citep{pocket_opening_enhanced,another_enhanced}.

\textbf{Deep Learning Approaches}. Molecular docking has been a field of interest in deep learning research during the last few years. The seminal work of DiffDock \citep{diffdock} introduced generative models to docking, achieving state-of-the-art performance. However, this work had already observed a lower performance in apo predicted structures compared to holo conformations. This finding prompted multiple studies aimed at generating improved protein conformations to use as inputs in computational workflows \citep{foldflow1,frameflow_im} or at directly modeling the joint distribution of protein and ligand conformations \citep{alphafold3}.

Closer to our work are ApolloDiff \footnote{We omit ApolloDiff as a baseline in our experiments due to the lack of weights and code availability.} \citep{apollodiff} and SBAlign \citep{sbalign}.  ApolloDiff focuses on the generation of holo conformations using an equivariant diffusion model conditioned on the apo conformation and sequence. SBAlign, meanwhile, formulates an alternative approach to solving the Schrödinger Bridge \citep{leonard2013survey} problem under the assumption that samples from the source and target distributions are available and paired and recovers a stochastic trajectory between them.

\section{Methods}

In this work we seek to learn a mapping between two explicit data distributions: the source distribution of apo structures, and the target distribution of holo structures. In this context, flow matching (FM) \citep{lipman2023flowmatchinggenerativemodeling} is a generative modeling paradigm which provides a simulation-free way of learning continuous normalizing flows (CNFs) that generate data over a learned vector field \citep{chenetal}. Given two distributions $q_0$ and $q_1$, one can then learn the CNF $\psi_t (x)$ that transports $q_0$ to $q_1$ by optimizing the conditional flow matching (CFM) objective:

\begin{equation}
    \mathcal{L}_{\text{CFM}} := 
        \mathbb{E}_{t, x_0 \sim q_0, x_1 \sim q_1}
        \left[ \left\| v_t(x) - u_t(x | x_1) \right\|^2 \right]
\end{equation}

where $u_t(x|x_1)$ is the conditional vector field associated with the interpolation between $x_0$ and $x_1$ and $v_t(x)$ can be parameterized with a neural network.

Previous works have extended FM to Riemannian Manifolds \citep{chen2024flowmatchinggeneralgeometries} and generalized it to use arbitrary source and target distributions \citep{tong2023improving, albergo2023stochastic}, postulating FM as a suitable framework for a generative model of holo structures given apo samples and allowing us to take advantage of several novel developments in the protein structure generation field \citep{foldflow1, alphaflow}.

Namely, we follow FoldFlow and parameterize backbones as  SE(3)-equivariant frames that represent rigid transformations $T = (r, x)\in SE(3)$ that consist of a rotation $r \in SO(3)$ and a translation $x\in \mathbb{R}^3$ \citep{alphafold2}. This formulation allows us to decompose the CFM process independently in $SO(3)$ and $\mathbb{R}^3$. In each space, we can define conditional flows using the geodesic (\ref{eq:geodesic}) and linear (\ref{eq:interpolation}) interpolants, respectively, and use their associated vector fields to optimize the following simplified objectives:



\begin{equation} \label{eq:geodesic}
    \mathcal{L}_{SO(3)} (\theta) = \mathbb{E}_{t, q(r_0, r_1)}
    \left\|  v_\theta(t, r_t) - u_t(r_t|r_0, r_1) \right\|_{SO(3)}^2 , \quad \text{with} \quad r_t = \exp_{r_0} \left( t \log_{r_0} (r_1) \right)
\end{equation}

\begin{equation} \label{eq:interpolation}
    \mathcal{L}_{\mathbb{R}^3} (\theta) = \mathbb{E}_{t, q(x_0, x_1)}
    \left\|  v_\theta(t, x_t) - u_t(x_t|x_0, x_1) \right\|^2, \quad \text{with} \quad x_t = (1-t)x_0 + tx_1
\end{equation}

We optimize these objectives with the loss function used by FoldFlow, and additionally find increased performance with the addition of the Frame Aligned Point Error loss from AlphaFold2 \citep{alphafold2}. Moreover, we also include the auxiliary losses from FrameFlow \citep{frameflow}, which encourage a better representation of the backbone atoms. Therefore, the final loss is formulated as: 

\begin{equation}
     \mathcal{L}_{total} = \mathcal{L}_{\textsc{FoldFlow}} + \mathcal{L}_{\text{FAPE}} + \mathcal{L}_{\text{aux}}
\end{equation}

Further details on FM and the used loss functions can be found in Appendix \ref{sec:exp_methods}.

\section{Experiments}

\subsection{Predicting large conformational changes}

We first evaluate our method in the dataset of overall protein motions of the D3PM database \citep{d3pm}, a dataset comprised of apo-holo pairs where the Root Mean Squared Deviation (RMSD) of the C$_\alpha$ carbon atoms between bound and unbound states is $>$\ 2.0\AA, which we call D3PM-Large (see Appendix \ref{app:datasets} for details).

In Table \ref{tab:d3pm_test_set} we report the statistics of the RMSD between the C$_\alpha$ carbon atoms of the generated structures and the holo conformations. Additionally, we also compute $\Delta RMSD = RMSD_{apo-holo} - RMSD_{gen-holo}$ to asses whether the generated structures are closer to the holo conformation than the respective apo one. Here, positive values are indicative of good performance \citep{apollodiff}.


 \begin{table}[ht]
    \caption{Conformational changes results in the D3PM-Large Test Set. RMSD between generated conformations and true holo structures. Results marked with an asterisk (*) were obtained from \citet{sbalign}.}
    \centering
    \label{tab:d3pm_test_set}
    \begin{tabular}{lcccccccccc}
        \toprule
         & \multicolumn{3}{c}{RMSD (\AA) $\downarrow$} 
         & \multicolumn{3}{c}{$\Delta$RMSD (\AA) $\uparrow$} 
         & \multicolumn{3}{c}{\% RMSD (\AA) $<$ $\tau$ $\uparrow$}   \\
        \cmidrule(lr){2-4} \cmidrule(lr){5-7} \cmidrule(lr){8-10}
        \textbf{Methods} 
         & \textbf{Med.} 
         & \textbf{Mean} 
         & \textbf{Std} 
         & \textbf{Med.} 
         & \textbf{Mean} 
         & \textbf{Std} 
         & $\tau=2$ 
         & $\tau=5$ 
         & $\tau=10$ \\
        \midrule
        EGNN*                 & 19.99 & 21.37 & 8.21 & --- & --- & --- & 1\% & 1\% & 3\% \\
        SBAlign & 3.67  & 4.82  & 3.93 & 1.30 & 1.92 & 2.59 & 0\% & 71\% & 93\% \\
       Sesame  & 2.87  & 3.65  & 2.95 & 2.15 & 3.11 & 4.26 & 38\% & 82\% & 96\% \\
        \bottomrule
    \end{tabular}
\end{table}

Sesame outperforms previous baselines, achieving 38\% of predictions with an RMSD $<$\ 2.0\AA \, with respect to the reference holo structure. Moreover, it also showcases better performance in generating structures that are more similar to the holo conformations than their apo counterparts, aligning more closely with our objectives. This distinction is crucial, since a generated conformation may be similar to the holo structure yet still remain closer to the apo state. 

While our model demonstrates strong performance in modeling large conformational changes, most ligand-induced conformational changes occur on a smaller scale, with 88\% of the cases having a RMSD between corresponding C$_\alpha$ atoms of less than 2\AA, and with 75\% of the cases where it falls below 1\AA \, \citep{frimurer2003ligand}.

\subsection{Predicting pocket conformational changes}

The task of predicting small conformational changes can prove to be more challenging due to the even greater scarcity of high-quality data, as these subtle changes are more difficult to capture \citep{x_ray_mh}. Although the D3PM database also has a dataset for pocket motions where the RMSD between C$_\alpha$ atoms is lower than 2\AA \, (see Appendix \ref{app:datasets-d3pm_pocket}), D3PM-Pocket only contains 865 apo-holo pairs, which represents a challenge to build generalizable predictive models. To overcome this, we construct a dataset of apo-holo pairs by generating apo conformations from holo structures after collapsing pockets using MD simulations. Further details in Appendix \ref{app:pdbbind-md}. 

For a more complete comparison with previous methods, we also retrain SBAlign on this new dataset. More details on the retraining procedure can be found in Appendix \ref{app:sbalign}. We additionally use the D3PM-Pocket dataset to evaluate the performance of both models in an established benchmark.

\begin{table}[ht]
    \centering
    \caption{Conformational changes results in the PDBBind-MD Test Set. RMSD between generated conformations and true holo structures. SBAlign was retrained in the PDBBind-MD dataset and inference was performed for the best model hyperparameters.}
    \label{tab:pdbbindmd_test_set}
    \begin{tabular}{lcccccccccc}
        \toprule
         & \multicolumn{3}{c}{RMSD (\AA) $\downarrow$} 
         & \multicolumn{3}{c}{$\Delta$RMSD (\AA) $\uparrow$} 
         & \multicolumn{3}{c}{\% RMSD (\AA) $<$ $\tau$ $\uparrow$} \\
        \cmidrule(lr){2-4} \cmidrule(lr){5-7} \cmidrule(lr){8-10}
        \textbf{Methods} 
         & \textbf{Med.} 
         & \textbf{Mean} 
         & \textbf{Std} 
         & \textbf{Med.} 
         & \textbf{Mean} 
         & \textbf{Std} 
         & $\tau=0.15$ 
         & $\tau=0.2$ 
         & $\tau=0.5$ \\
        \midrule
        SBAlign  & 0.52  & 0.55  & 0.18 & 0.05 & 0.03 & 0.11 & 0 \% & 0 \% & 24.46 \% \\
        Sesame  & 0.18  & 0.18  & 0.03 & 0.39 & 0.39 & 0.03 & 7.57 \% & 82.16 \% & 100\% \\
        \bottomrule
    \end{tabular}
\end{table}

\begin{table}[H]
    \centering
    \caption{Conformational changes results in the D3PM-Pocket set. RMSD between generated conformations and true holo structures. SBAlign was retrained in the PDBBind-MD dataset and inference was performed for the best model hyperparameters.}
    \label{tab:d3pm_pocket}
    \begin{tabular}{lcccccccccc}
        \toprule
         & \multicolumn{3}{c}{RMSD (\AA) $\downarrow$} 
         & \multicolumn{3}{c}{$\Delta$RMSD (\AA) $\uparrow$} 
         & \multicolumn{3}{c}{\% RMSD (\AA) $<$ $\tau$ $\uparrow$} \\
        \cmidrule(lr){2-4} \cmidrule(lr){5-7} \cmidrule(lr){8-10}
        \textbf{Methods} 
         & \textbf{Med.} 
         & \textbf{Mean} 
         & \textbf{Std} 
         & \textbf{Med.} 
         & \textbf{Mean} 
         & \textbf{Std} 
         & $\tau=0.15$ 
         & $\tau=0.2$ 
         & $\tau=0.5$ \\
        \midrule
        SBAlign  & 0.77  & 1.21  & 1.08 & 0.001 & 0.001 & 0.002 & 1.16 \% & 2.33 \% & 18.60 \% \\
        Sesame  & 0.49  & 0.91  & 1.22 & 0.18 & 0.19 & 0.1 & 5.59 \% & 13.61 \% & 50.67 \% \\
        \bottomrule
    \end{tabular}
\end{table}

As we can observe on Tables \ref{tab:pdbbindmd_test_set} and \ref{tab:d3pm_pocket}, we outperform the existing baseline in $\Delta$RMSD both in the PDBBind-MD and D3PM-Pocket datasets, highlighting the potential of Sesame in capturing small conformational changes.

Summary statistics for each dataset are provided in Appendix \ref{app:dataset-stats} to better contextualize the reported results. Additionally, we report results for cross-inferences, where we evaluate the models trained on large conformational changes on small ones and vice versa to assess generalization capabilities across different data distributions.

\subsection{Sesame Facilitates Cryptic Pocket Identification and Characterization}
Due to the increasing interest in drug discovery in targeting cryptic and transient pockets, multiple algorithms for identifying and characterizing them have been designed. One example is PocketMiner, a recent deep learning method which predicts the probability of a residue belonging to a cryptic site \citep{meller2022predicting}. Testing Sesame alongside pocket prediction algorithms can help assess a potential synergy, since Sesame is trained to generate proteins closer to holo states from apo structures. This could improve the prediction of cryptic and transient pockets and allow for better identification of the residues involved in them. Additionally, if the relevant residues are detected more effectively, it would further confirm that the generated conformations are more holo-like.

To evaluate how Sesame could aid in this endeavor, we generate holo-like structures for the validation and test sets used in \citet{meller2022predicting}, a set of known cryptic pockets, labeling as positive those residues closer than 8 \AA \, to the ligand. We then run PocketMiner on the apo, holo and predicted conformations and obtain the precision-recall and receiver operating characteristic (ROC) curves for each. 

\begin{figure}[ht]
    \centering
    \begin{subfigure}{.5\textwidth}
      \centering
      \includegraphics[width=\linewidth]{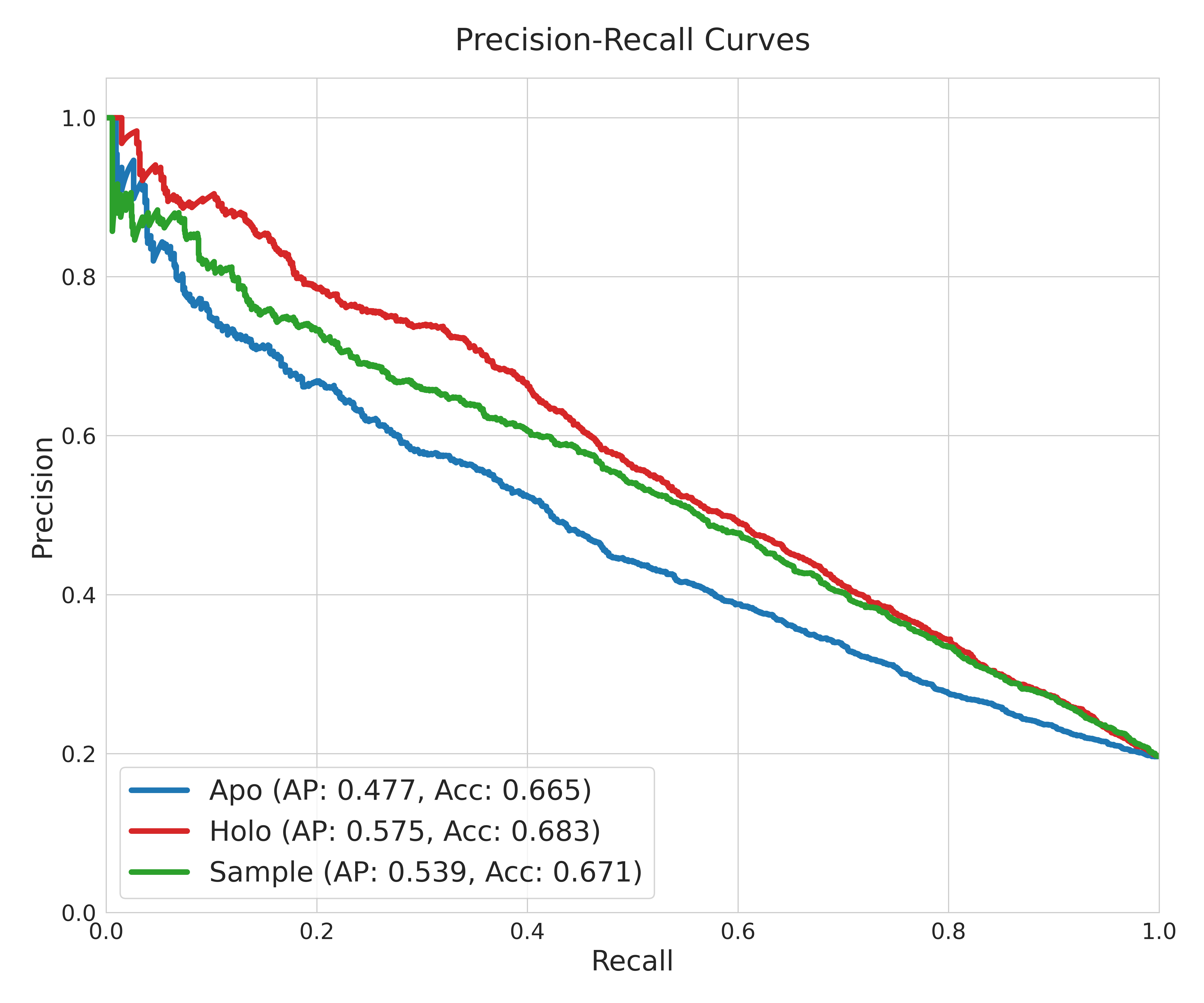}
      \label{fig:pminer_pr}
    \end{subfigure}%
    \begin{subfigure}{.5\textwidth}
      \centering
      \includegraphics[width=\linewidth]{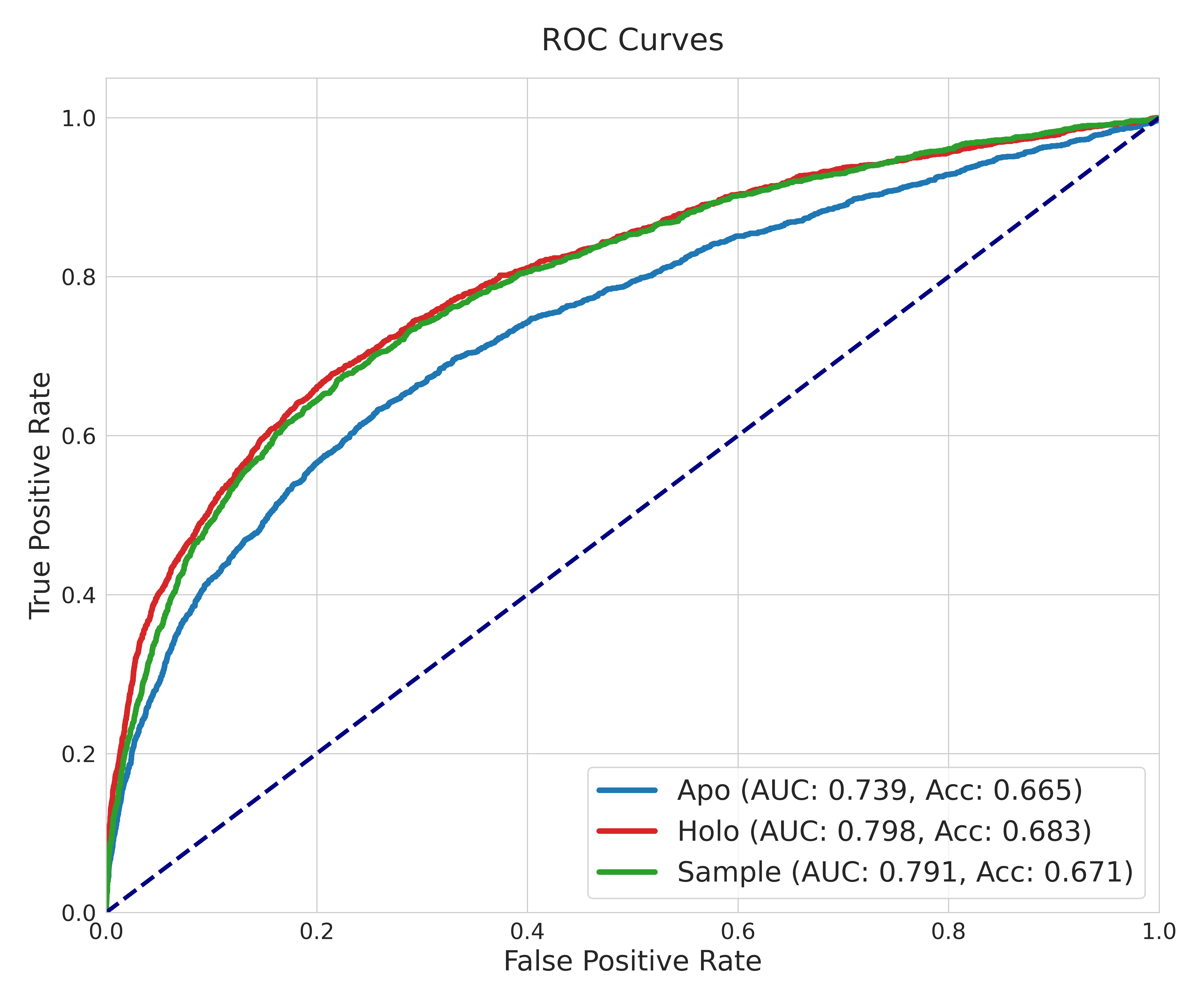}
      \label{fig:pminer_roc}
    \end{subfigure}
    \caption{Precision-Recall (left) and ROC (right) curves for PocketMiner predictions.}
    \label{fig:pminer_predictions}
\end{figure}

In Figure \ref{fig:pminer_predictions}, we can observe an increase in performance in PocketMiner predictions on Sesame's generated structures versus their apo counterparts. This can indicate that the generated conformations are more geometrically similar to holo structures, increasing the reliability of pocket detection algorithms. For further discussion and visualizations of results obtained see Appendix \ref{app:viz}.


%

\subsection{Sesame generates viable structures for molecular docking}


In this section we seek to evaluate the performance of Sesame generated structures using them as input for a standard docking protocol. A common way of measuring a docking software's performance is reproducing a crystallographically resolved structure, with the degree of geometric similarity serving as said measurement of the performance \citep{self_docking}. In this context, we can evaluate the ability of Sesame generated structures to accept compound geometries that resemble those in the crystal holo conformations.

It is important to note that Sesame employs only backbone atoms, meaning side chains are not explicitly determined. Since most docking software requires side chains as input, and they are vital for determining the ligand's correct position, their positions were predicted using Prime \citep{prime}. To ensure a fair comparison, the side chains of both holo and apo structures were also reconstructed following the same protocol and included as baselines. Docking performance was then evaluated using RMSD, with all calculations performed against the ligand structure in the holo X-ray.

The two PDB complexes were chosen based on the availability of both apo and holo structures with high resolution and their clinical significance \citep{clinical1,clinical2}.

\begin{table}[h]
    \centering
    \caption{RMSD and corresponding docking poses for the Holo versus Reconstructed Holo, Apo and Model. (PDBs: 4ZZI, 4XKQ)  }\label{tab:4ZZI}
    
    \begin{tabular}{lcc}
        \toprule
        & RMSD (\AA) $\downarrow$ & Docking Pose\\
        \midrule
        Holo X-ray vs Reconstructed Holo & 1.622  & Figure \hyperref[fig:docking]{\ref*{fig:docking}.A}\\
        Holo X-ray vs Reconstructed Apo & 5.704   & Figure \hyperref[fig:docking]{\ref*{fig:docking}.B}\\
        Holo X-ray vs Model & 3.724     & Figure \hyperref[fig:docking]{\ref*{fig:docking}.C}\\
        \bottomrule
    \end{tabular}

\end{table}

In the results corresponding to the complex 4ZZI in Table \ref{tab:4ZZI}, we obtain an increased performance when compared to the apo structure, indicating that the obtained backbone structure is better suited for docking protocols. The low RMSD in the X-ray versus reconstruction comparison confirms the validity of the side chain prediction protocol. Further discussion and additional results of the 4LVT complex can be found in Appendix \ref{app:sidechsesame}.

For full details of the corresponding docking methods, protocol and visualization of the obtained poses that were used for the RMSD calculations see Appendix \ref{app:dock} and Appendix \ref{app:vizu_docking}, respectively.

\section{Conclusions and Future Work}

In this work, we introduce Sesame, a novel generative model that leverages the flow matching framework to construct a generative process between two data distributions, learning how to transform apo-like protein conformations into holo-like ones. Our model surpasses existing baselines in capturing both large and small conformational changes. Furthermore, we demonstrate that the generated conformations aid in identifying cryptic binding pockets and serve as effective input structures for molecular docking, leading to improved results compared to their original apo counterparts. These findings indicate that Sesame offers a scalable and efficient approach for enhancing virtual screening workflows and streamlining the drug discovery pipeline.

Future work aims to address the main limitations of the model. Namely, this includes extending the framework to incorporate side chain modeling, as they play a fundamental role in protein-ligand interactions, binding affinity and docking accuracy. Additionally, we plan to expand our data generation efforts using molecular dynamics simulations that capture transitions across energetic barriers, thereby increasing the diversity of targetable binding pockets and providing richer datasets that encompass both subtle and large-scale movements and allow for an end-to-end model capable of accurately modeling both.

\bibliography{iclr2025_conference}
\bibliographystyle{iclr2025_conference}
\newpage
\appendix

\section{Datasets} \label{app:datasets}

\subsection{D3PM Database} \label{app:datasets-d3pm}


The D3PM database collects all kinds of protein motions, including overall structural changes upon ligand binding and the inherently flexibility of protein, which we refer to as D3PM-Large, as well as flap movements of residues within binding pocket; which we do so as D3PM-Pocket \citep{d3pm}.

\subsubsection{D3PM-Large} \label{app:datasets-d3pm_large}

The initial dataset is sourced from the D3PM-Large, where the RMSD of the C$_\alpha$ carbon atoms between bound and unbound states is $>$\ 2.0\AA, following the protocol outlined in SBAlign \citep{sbalign}. First, apo-holo pairs are selected and filtered to ensure that the reported RMSD of $C_\alpha$ exceeds 3Å. The selected pairs are then aligned using the Kabsch \citep{kabsch} algorithm, and the RMSD of these $C_\alpha$ is recalculated. Pairs are retained if the recalculated RMSD falls within a small margin of error, indicating minimal reconstruction error. Lastly, proteins with more than one chain are removed from the set. Then the structures are randomly split in training, validation and tests sets.

\begin{table}[h]
    \centering
    \caption{D3PM-Large training, validation, and test sets.}
    
    \begin{tabular}{lcccc}
        \toprule
        &  \# Unique chains & \# Apo chains & \# Holo chains & \# Unique Proteins \\
        \midrule
        Train & 1291 & 799 & 1133 & 443 \\
        Val   & 150  & 136  & 147   & 73  \\
        Test  & 150  & 141  & 144   & 74  \\
        \bottomrule
    \end{tabular}
    
\end{table}

\subsubsection{D3PM-Pocket} \label{app:datasets-d3pm_pocket}

This dataset contains pairs of apo-holo proteins where the overall RMSD between C$_\alpha$ atoms is $<$\ 2.0\AA. We use the entire dataset for evaluation, with no other data processing than the one outlined in \ref{app:data_proc}

\begin{table}[h]
    \centering
    \caption{D3PM-Pocket training, validation, and test sets.}
    
    \begin{tabular}{lcccc}
        \toprule
        &  \# Unique chains & \# Apo chains & \# Holo chains & \# Unique Proteins \\
        \midrule
        Test  & 1723  & 861 & 862   & 848  \\
        \bottomrule
    \end{tabular}
    
\end{table}

\subsection{PDBBind-MD}\label{app:pdbbind-md}
However informative, the D3PM-Pocket dataset has quite a small size. Due to the lack of additional holo-apo pairing data, we decided to artificially generate more data using molecular dynamics. While it is quite complicated to generate holo conformations from apo structures, the reverse procedure is a simpler problem. In this dataset we take advantage of the PDBBind dataset \citep{pdbbind} split in temporal splits following DiffDock \citep{diffdock}. This set contains bound holo structures, and employing molecular dynamics, we collapse the pockets to generate an artificial apo structure, that is naturally paired with the original holo one. 

First a quality filter is passed on the protein structures, which consists in the removal of structures with backbone breaks.
Afterwards, for each remaining entry, the ligand was removed, and the protein structure was prepared for MD simulations. The proteins were parameterized using the AMBER ff14SB \citep{ambertools,ff14SB} force field in the gas phase to ensure consistency of force field parameters across all systems.

Each system underwent an initial heating phase followed by an equilibration before trajectory analysis. The heating and equilibration steps were conducted using implicit solvent conditions, with the generalized Born model, to minimize computational cost while maintaining biophysical relevance. 

All systems were initially heated from 0 K to 550 K over 1000 steps. This heating phase utilized Langevin dynamics (ntt=3) with a collision frequency of 100.0 $ps^{-1}$ to maintain thermal stability. The timestep was set to 0.0005 ps, ensuring controlled energy fluctuations during heating.
After heating, each system was equilibrated for 4 ps at a constant temperature of 550 K. A higher Langevin collision frequency (200.0 $ps^{-1}$) was used during equilibration to stabilize the system at 550 K while preventing excessive energy fluctuations. The timestep was increased to 0.001 ps to allow for a more efficient sampling of the conformational space.

To analyze the stability and dynamic behavior of each protein, residues within a 5\AA\, radius of the removed ligand were extracted. The RMSD of these residues was calculated throughout the MD trajectory to evaluate conformational changes in the binding pocket. This protocol was used to determine which systems exhibited the most significant conformational changes in the binding site after ligand removal. Afterwards, proteins having a minimum RMSD change of 0.5\AA \,were kept.

\begin{table}[H]
    \centering
    \caption{PDBBind-MD training, validation, and test sets.}
    
    \begin{tabular}{lcccc}
        \toprule
        & \# Unique Chains & \# Apo chains & \# Holo chains & \# Unique Proteins \\
        \midrule
        Train & 6596 & 6596 & 6596 & 2334 \\
        Val   & 395  & 395  & 395 & 353  \\
        Test  & 185  & 185  & 185 & 80  \\
        \bottomrule
    \end{tabular}
    
\end{table}

\subsection{Data Processing} \label{app:data_proc}

Apo and holo chains are featurized following AlphaFold2 \citep{alphafold2}, where each residue is associated with SE(3)-equivariant frames that map rigid transformations onto an idealized backbone configuration. Each frame is defined by the atomic coordinates of the nitrogen (N), alpha carbon (C$_\alpha$), carbonyl carbon (C), and oxygen (O) atoms, with transformations incorporating experimental bond angles and lengths \citep{engh2012structure}. The backbone oxygen position is determined by an additional rotation around the C$_\alpha$-C bond. The final representation encodes all heavy atom coordinates as structured transformations in \(\mathbb{R}^{N \times 4 \times 3}\), ensuring consistency across conformational states. 

In this step, we additionally add a mask in the holo structure for pocket residues, defined as those within a 8 \AA \, distance cutoff of the ligand. After this, apo-holo pairs undergo local sequence alignment and only matching positions are kept. Then apo structures are centered by substracting the center of mass from the C$_\alpha$ positions and holo structures are superimposed onto them using Kabsch alignment of the C$_\alpha$ atoms. During apo-holo alignment, we transfer the holo-defined pocket mask to the apo and retain up to 512 residues, cropping the excess when necessary.

\subsection{Dataset Statistics} \label{app:dataset-stats}

We report summary statistics for each dataset, as they help interpret results, putting in context the obtained metrics.

\begin{table}[ht]
   \centering
   \caption{Summary statistics for the datasets.}
   \label{tab:dataset_statistics}
   \begin{tabular}{lcccccccccc}
       \toprule
        & \multicolumn{3}{c}{Length} 
        & \multicolumn{3}{c}{C$_\alpha$ RMSD} 
        & \multicolumn{3}{c}{Seq. Id. (\%)} \\
       \cmidrule(lr){2-4} \cmidrule(lr){5-7} \cmidrule(lr){8-10}
       \textbf{Dataset} 
        & \textbf{Med.} 
        & \textbf{Mean} 
        & \textbf{Std} 
        & \textbf{Med.} 
        & \textbf{Mean} 
        & \textbf{Std} 
        & \textbf{Med.} 
        & \textbf{Mean} 
        & \textbf{Std}   \\
       \midrule
       PDBBind-MD  & 223  & 244  & 127.48 & 0.58  & 0.61  & 0.31 & 100.00  & 99.98  & 1.35 \\
       D3PM-Pocket & 267 & 290 & 121.49 & 0.72 & 1.10 & 1.19 & 97.67 & 94.70 & 10.28 \\
       D3PM-Large  & 270 & 285 & 136.35 & 3.07  & 4.55  & 4.27 & 95.20 & 87.02 & 19.88 \\
       \bottomrule
   \end{tabular}
\end{table}

\section{Training SBAlign} \label{app:sbalign}

To retrain the SBAlign model on the PDBBind-MD dataset, we modified some hyperparameters to better suit the model to the task of predicting small conformational changes. Primarily, we use a lower diffusivity value, which controls the variance of the stochastic process. 

We hypothesize that this led to some instabilities in the loss function, defined as:

\begin{equation}
L(\theta, \phi) := \mathbb{E} \left[ \int_0^1 \left\| \mathds{G} - \left(b_t^\theta + m^\phi(X_t)\right) \right\|^2 + \lambda_t \|m^\phi(X_t)\|^2 \, dt \right],
\end{equation}

with $\mathds{G}$:

\begin{equation}\label{eq:boom}
\frac{\mathbf{x}_1 - X_t}{\beta_1 - \beta_t}
\end{equation}

where \(\beta_t\) is given by:

\begin{equation}
\beta_t := \int_0^t g_s^2 \, ds.
\end{equation}

As \(g_s^2 \to 0\), \(\beta_t \to 0\), and consequently, the term in Equation \ref{eq:boom} grows larger, causing large gradients. These gradients can destabilize the optimization, leading to divergence and therefore to increasingly larger loss values during the training. This hypothesis aligns with the observed instability when training with low diffusivity.

We also note this affects inference \footnote{In all inferences for SBAlign across this work, we generate 10 structures in 10 sampling steps and report metrics for the best one.} equally. In Table \ref{tab:pdbbindmd_test_set_bad} we report metrics for several values of $g$ with the model trained on PDBBind-MD, and in Table \ref{tab:d3pm_test_set_bad} we present similar results for the original SBAlign model. These results indicate that SBAlign is very sensitive to this parameter and struggles to generalize across a broader range of motions.

\begin{table}[H]
    \centering
    \caption{Conformational changes results in the PDBBind-MD Test Set. RMSD between generated conformations and true holo structures after doing inference using SBAlign at different $g$ values.}
    \label{tab:pdbbindmd_test_set_bad}
    \begin{tabular}{lccccccccccc}
        \toprule
         & \multicolumn{3}{c}{RMSD (\AA) $\downarrow$} 
         & \multicolumn{3}{c}{$\Delta$RMSD (\AA) $\uparrow$}  \\
        \cmidrule(lr){2-4} \cmidrule(lr){5-7} \cmidrule(lr){8-10}
        \textbf{$g$ value} 
         & \textbf{Median} 
         & \textbf{Mean} 
         & \textbf{Std} 
         & \textbf{Median} 
         & \textbf{Mean} 
         & \textbf{Std}\\
        \midrule
        0.001  & 0.57  & 0.59  & 0.15 & 0.0009 & 0.0009 & 0.0003 \\
        0.01   & 0.52  & 0.55  & 0.18 & 0.05 & 0.03 & 0.1 \\
        0.02  & 0.77  & 0.87  & 0.32 & -0.28 & -0.21 & 0.28 \\
        \bottomrule
    \end{tabular}
\end{table}

\begin{table}[H]
    \centering
    \caption{Conformational changes results in the D3PM-Large Test Set. RMSD between generated conformations and true holo structures after doing inference using SBAlign at different $g$ values.}
    \label{tab:d3pm_test_set_bad}
    \begin{tabular}{lccccccccccc}
        \toprule
         & \multicolumn{3}{c}{RMSD (\AA) $\downarrow$} 
         & \multicolumn{3}{c}{$\Delta$RMSD (\AA) $\uparrow$}  \\
        \cmidrule(lr){2-4} \cmidrule(lr){5-7} \cmidrule(lr){8-10}
        \textbf{$g$ value} 
         & \textbf{Median} 
         & \textbf{Mean} 
         & \textbf{Std} 
         & \textbf{Median} 
         & \textbf{Mean} 
         & \textbf{Std}\\
        \midrule
        0.1  & 4.43  & 6.70  & 5.16 & 0.036 & 0.029 & 0.003 \\
        1.0   & 3.80  & 4.98  & 3.95 & 1.18 & 1.76 & 2.52  \\
        2.0  & 7.69  & 8.87  & 4.14 & -2.60 & -2.11 & 3.06 \\
        \bottomrule
    \end{tabular}
\end{table}

\newpage

\section{Cross-Inferences Results}

In this section, we assess the capabilities of each model to generalize
across different data distributions. To this end, we perform inferences with the models trained
on large conformational changes to predict small ones and vice versa. 

As we can see in Table \ref{tab:cross_inferences}, both models suffer from reduced performance when predicting conformational changes of a magnitude different that those it was trained on. Intuitively, we observe that the models trained on large conformational changes fail by producing movements that are too large and even lead corrupt structures, as evidenced by RMSD values much higher than the range found in the apo-holo pairs in the dataset. Conversely, the ones trained on small ones are able to capture only a small part of movement and fail to capture the full range of it, reflected in the combination of low $\Delta$RMSD values and high RMSD ones, but that fall within the distribution of apo-holo pairs in the dataset.

While these results are perhaps to be expected, it highlights a key limitation of the models, which ultimately stems from a lack of diverse, high-quality data.

\begin{table}[ht]
    \centering
    \caption{Conformational changes results for cross-inferences. RMSD between generated conformations and true holo structures after doing inference in the test sets of the corresponding datasets.}
    \label{tab:cross_inferences}
    \begin{tabular}{l cc ccc ccc}
        \toprule
        & \multicolumn{2}{c}{Dataset Used} 
        & \multicolumn{3}{c}{RMSD (\AA) $\downarrow$} 
        & \multicolumn{3}{c}{$\Delta$RMSD (\AA) $\uparrow$}\\
        \cmidrule(lr){2-3} \cmidrule(lr){4-6} \cmidrule(lr){7-9}
        \textbf{Model} & \textbf{Train} & \textbf{Inference} 
         & \textbf{Med.} & \textbf{Mean} & \textbf{Std} 
         & \textbf{Med.} & \textbf{Mean} & \textbf{Std} \\
        \midrule
        SBAlign  & D3PM-Large & PDBBind-MD  & 2.23  & 2.36  & 0.45 & -1.66 & -1.77 & 0.45 \\
        Sesame & D3PM-Large & PDBBind-MD  & 2.38  & 2.50  & 0.90 & -1.79 & -1.93 & 0.90 \\
        SBAlign  & PDBind-MD & D3PM-Large  & 4.46  & 6.74  & 5.16 & 0.025  & 0.003  & 0.31  \\
        Sesame  & PDBind-MD & D3PM-Large  & 4.22  & 6.31  & 5.04 & 0.12  & 0.12  & 0.16  \\
        \bottomrule
    \end{tabular}
\end{table}

\section{Visualizations of Cryptic Pocket Prediction} \label{app:viz}

Here, we provide two example of PocketMiner's predictions for apo, holo and generated structures, showing comparisons of the structures and PocketMiner's per-residue predictions, where residues predicted to belong to a cryptic binding pocket with a probability $p > 0.5$ are highlighted in red.

\begin{figure}[H]
    \centering
    \begin{subfigure}{.4\textwidth}
      \centering
      \includegraphics[width=0.95\linewidth]{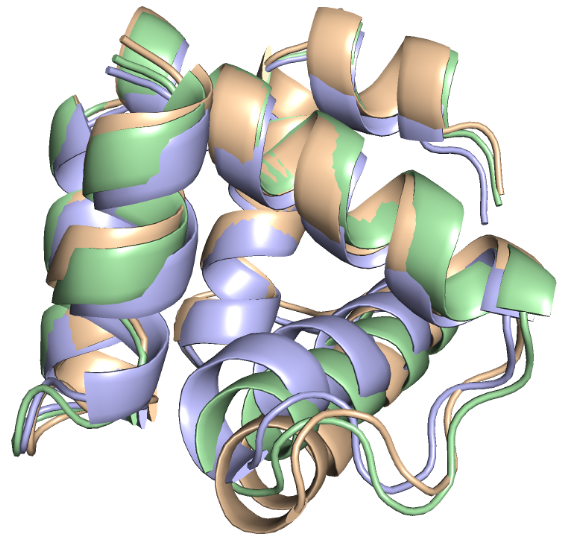}
      \label{fig:pminer_1kx9_prot}
    \end{subfigure}%
    \begin{subfigure}{.55\textwidth}
      \centering
      \includegraphics[width=\linewidth]{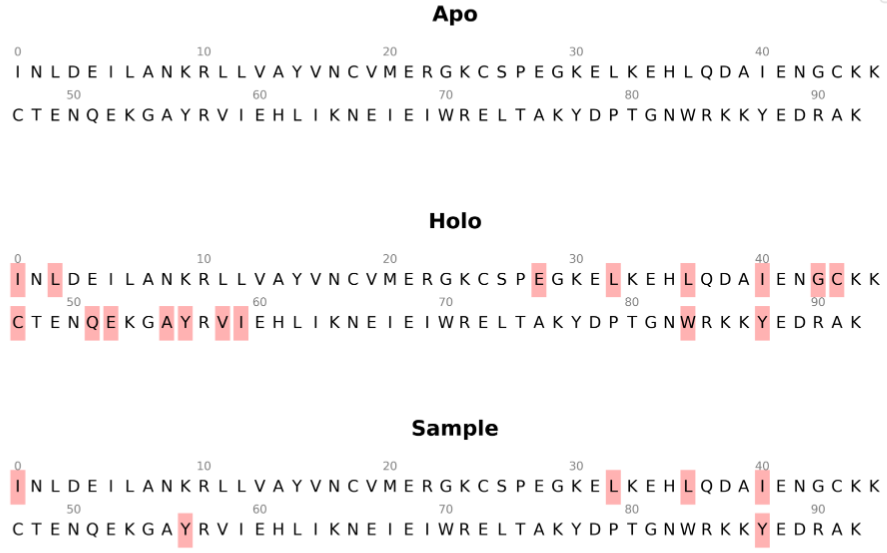}
      \label{fig:pminer_1kx9_seq}
    \end{subfigure}
    \caption{Left: Apo (blue), Holo (yellow) and Generated Sample (green). Right: PocketMiner per-residue predictions for each structure (PDBs: 1KX9-B, 1N8V-B)}
    \label{fig:pminer_1kx9}
\end{figure}

\begin{figure}[H]
    \centering
    \begin{subfigure}{.5\textwidth}
      \centering
      \includegraphics[width=\linewidth]{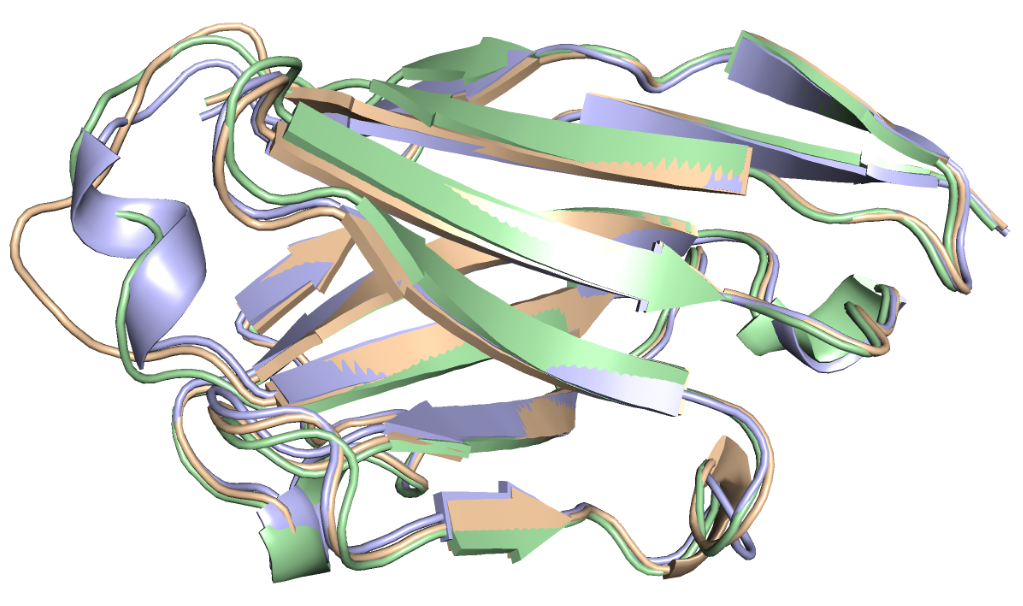}
      \label{fig:pminer_3qxw_prot}
    \end{subfigure}%
    \begin{subfigure}{.5\textwidth}
      \centering
      \includegraphics[width=\linewidth]{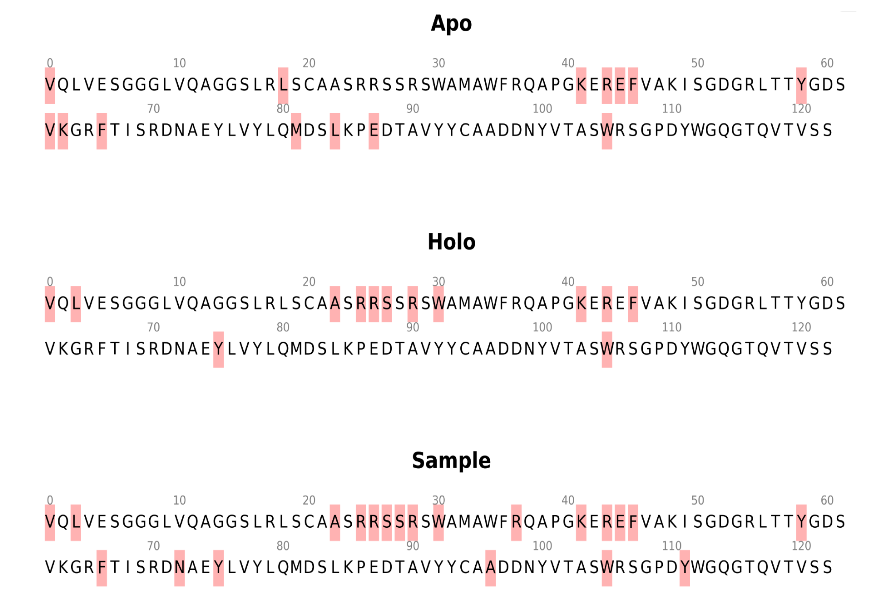}
      \label{fig:pminer_3qxw_seq}
    \end{subfigure}
    \caption{Left: Apo (blue), Holo (yellow) and Generated Sample (green). Right: PocketMiner per-residue predictions for each structure (PDBs: 3QXW-B, 3QVX-A)}
    \label{fig:pminer_3qxw}
\end{figure}

As seen in Figures \ref{fig:pminer_1kx9} and \ref{fig:pminer_3qxw}, the generated structures are closer to the holo conformations for some regions of the proteins, and PocketMiner's per-residue predictions are more closely aligned to those of the holo than the apo as well. 

\section{Docking Methods and Visualizations}

\subsection{Methods}
\label{app:dock}
Protein Data Bank (PDB) was the source for the apo and holo X-ray structures. To ensure a fair comparison with the model, side chains (except beta carbon), water, ions, and heteroatom molecules were removed from the crystal structures. Missing side chains in both X-rays and models were reincorporated using Prime \citep{prime}. The Protein Preparation Wizard \citep{prot_prep_wiz} in Schrödinger (Schrödinger Release 2024-4) was used to add hydrogens and assign bond orders. Hydrogen bond and side-chain orientations were optimized at  pH 7.4 with PROPKA \citep{propka}. Finally, in those cases where side-chains were reincorporated they were minimized with Prime, using OPLS4 force field \citep{opls4} and solvation VSGB \citep{vsgb}. Otherwise, we applied a restrained minimization of 0.3\AA\, RMSD. Grid-based receptor preparation was performed by centering in the ligand-binding site identified from the X-ray structure. Ligand docking was conducted using the standard Glide SP \citep{glide_sp} protocol. Figures were generated with PyMOL \citep{pymol}.

\subsection{Sesame helps correct side chain prediction errors}
\label{app:sidechsesame}

We further evaluate Sesame using the 4LVT complex. Interestingly, in this case, our side chain reconstruction protocol performs significantly worse. Comparing the RMSD results of the reconstructed side chains to those obtained from self-docking using the X-ray side chains, we observe a substantially lower geometric similarity. We hypothesize that this is due to the apo conformation having a more closed pocket than the holo conformation, leading to a worse placement of side chains.

However, applying the same protocol to Sesame's generated structure yields improved results. This further reinforces the potential of extending the method to include side chains as a promising direction for future research.

\begin{table}[h]
    \centering
    \caption{RMSD and corresponding docking poses for the X-ray Holo versus X-ray Holo, Reconstructed Holo, X-ray Apo and Model. (PDBs: 4LVT, 1GJH)}
    \label{tab:4LVT}
    
    \begin{tabular}{lcc}
        \toprule
        & RMSD (\AA) $\downarrow$ & Docking Pose\\
        \midrule
        Holo X-ray vs Holo X-Ray & 3.063  & Figure \hyperref[fig:docking]{\ref*{fig:docking}.D}\\
        Holo X-ray vs Reconstructed Holo & 8.752   & Figure \hyperref[fig:docking]{\ref*{fig:docking}.E}\\
        Holo X-ray vs Apo X-Ray & 14.886  & Figure \hyperref[fig:docking]{\ref*{fig:docking}.F}\\
        Holo X-ray vs Model & 6.081    & Figure \hyperref[fig:docking]{\ref*{fig:docking}.G}\\
        
        \bottomrule
    \end{tabular}
    
\end{table}

\subsection{Visualizations}
\label{app:vizu_docking}

\begin{figure}[htbp]
    \centering
    \includegraphics[width=\linewidth]{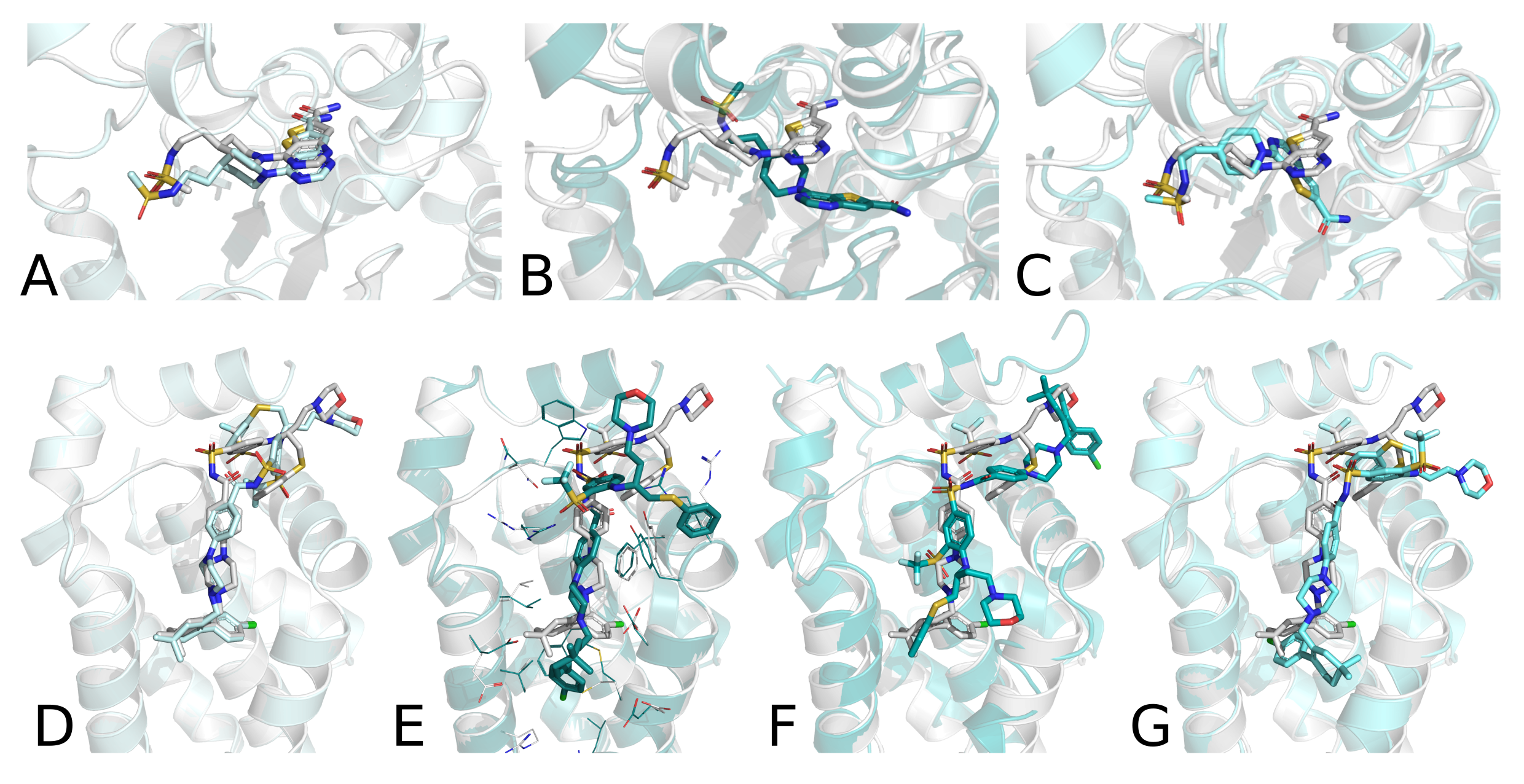}
    \caption{Docking Poses used for RMSD calculations in Tables \ref{tab:4ZZI} and \ref{tab:4LVT}. Alphabetical ordering follows order of appearance in the respective tables.}
    \label{fig:docking}
\end{figure}

Figure \ref{fig:docking} displays the highest-ranked docking pose for each structure based on Glide score, which was subsequently used for RMSD calculations.

\section{Detailed Methods}
\label{sec:exp_methods}

In this section, we aim to provide a more detailed theoretical background on flow matching and in the used loss functions in our work.

\textbf{Flow Matching}. Flow matching is generative modeling paradigm that was introduced as a efficient and simulation-free way to learn continuous normalizing flows (CNFs) \citep{lipman2023flowmatchinggenerativemodeling}. CNFs are a class of deep generative models that generate data by integrating an ordinary differential equation over a learned vector field \citep{chenetal}.

Let $q_0$ and $q_1$ be two distributions, one can learn a vector field $v_t$ that induces a CNF $\psi_t(x)$ that transports $q_0$ to $q_1$:

\begin{equation}
    q_1(x) = [\psi_1]_\#q_0(x)
\end{equation}

where $\#$ denotes the pushforward operator. 

The marginal vector field $v_t(x)$ can then be parametrized with a neural network $v_t(x_t; \theta)$ and learned by defining conditional flows $\psi_t(x_0 |x_1)$ interpolating between $x_0 \sim q_0$ and $x_1 \sim q_1$ and their associated vector field $u_t(x_t|x_1)$, and regressing against the conditional vector field with the CFM objective:

\begin{equation}
    \mathcal{L}_{\text{CFM}} := 
        \mathbb{E}_{t, x_0 \sim q_0, x_1 \sim q_1}
        \left[ \left\| v_t(x) - u_t(x | x_1) \right\|^2 \right],
\end{equation}

Although originally constrained to a Gaussian starting distribution, flow matching was generalized, relaxing said constraint and allowing arbitrary couplings by taking a joint distribution $q(z)$, over several possible choices \citep{tong2023improving, albergo2023stochastic}.

Flow matching has additionally been extended to Riemannian Manifolds \citep{chen2024flowmatchinggeneralgeometries}. On a manifold $\mathcal{M}$, the CNF $\psi_t(x)$ can be defined by integrating along a time-dependent vector field $v_t(x) \in \mathcal{T}_x\mathcal{M}$, where $\mathcal{T}_x\mathcal{M}$ is the tangent space of the manifold at $x \in \mathcal{M}$.

As such, we can similarly learn the marginal vector field with a similar objective:

\begin{equation}
    \mathcal{L}_{\text{CFM}} := 
        \mathbb{E}_{t, x_0 \sim q_0, x_1 \sim q_1}
        \left[ \left\| v_t(x) - u_t(x | x_1) \right\|^2_g \right],
\end{equation}

Where $||\cdot||^2_g$ is the norm induced by the Riemannian metric $g$, and the conditional flows $\psi_t(x)= \psi_t(x_0 | x_1)$ can be defined by interpolating along the geodesic paths for simple manifolds \citep{frameflow}.

The extension of flow matching to Riemannian Manifolds makes this framework readily applicable to the protein modeling field and to our goal of a generative model of holo structures given apo ones. More specifically, we follow FoldFlow and parameterize backbones as  SE(3)-equivariant frames that represent rigid transformations $T = (r, x)\in SE(3)$ that consist of a rotation $r \in SO(3)$ and a translation $x\in \mathbb{R}^3$ \citep{alphafold2}. This formulation allows us to decompose the CFM process independently in $SO(3)$ and $\mathbb{R}^3$. In each space, we can define conditional flows using the geodesic (\ref{eq:geodesic}) and linear (\ref{eq:interpolation}) interpolants, respectively, and use their associated vector fields to optimize the following simplified objectives:

\begin{equation} \label{eq:geodesic2}
    \mathcal{L}_{SO(3)} (\theta) = \mathbb{E}_{t, q(r_0, r_1)}
    \left\|  v_\theta(t, r_t) - u_t(r_t|r_0, r_1) \right\|_{SO(3)}^2 , \quad \text{with} \quad r_t = \exp_{r_0} \left( t \log_{r_0} (r_1) \right)
\end{equation}

\begin{equation} \label{eq:interpolation2}
    \mathcal{L}_{\mathbb{R}^3} (\theta) = \mathbb{E}_{t, q(x_0, x_1)}
    \left\|  v_\theta(t, x_t) - u_t(x_t|x_0, x_1) \right\|^2, \quad \text{with} \quad x_t = (1-t)x_0 + tx_1
\end{equation}

Where the conditional vector fields for rotation and translation take the following expressions, respectively:

\begin{equation}
    u_t(x_t|x_0, x_1) = x_1 - x_0 
\end{equation}

\begin{equation}
    u_t(r_t|r_0, r_1) = \log_{r_t}(r_0)/t
\end{equation}

Following previous works \citep{frameflow}, we perform all modelling within the zero center of mass (CoM) subspace, which guarantees that the distribution of the sampled frames is SE(3)-invariant, and pre-align with the Kabsch algorithm to remove global rotations that result in increase kinetic enery in the ODE, and parameterize the velocity predictions $v_{\theta}(t, x_t)$ as a prediction of the starting point $\hat{x}_0$ given $x_t$ \citep{frameflow, foldflow1, alphaflow}.

\textbf{Final loss function}. Finally, the complete loss function is formulated as: 

\begin{equation}
     \mathcal{L} = \mathcal{L}_{\text{FOLDFLOW}} + \mathcal{L}_{\text{FAPE}} + \mathds{1}_{\{t < 0.25\}} \lambda_{\text{aux}} \mathcal{L}_{\text{aux}} ,
\end{equation}

with 

\begin{equation}
    \mathcal{L}_{\text{FOLDFLOW}} = \mathcal{L}_{\text{FOLDFLOW}-SO(3)} + \mathcal{L}_{\text{FOLDFLOW}-\mathbb{R}^3},
\end{equation}

and:

\begin{equation}
    \mathcal{L}_{\text{FOLDFLOW}-\mathbb{R}^3} = \mathbb{E}_{t\sim \mathcal{U}([0,1]), q(x_0,x_1), \rho_t (x_t | x_0, x_1)} \left\|  v_\theta(t, x_t) - (x_1 - x_0) \right\|^2
\end{equation}

\begin{equation}
    \mathcal{L}_{\text{FOLDFLOW}-SO(3)} = \mathbb{E}_{t \sim \mathcal{U}([0,1]), q(r_0, r_1), \rho_t (r_t | r_0, r_1)} \left\|  v_\theta(t, r_t) - \log_{r_t}(r_0)/t \right\|^2_{SO(3)}
\end{equation}

\textbf{FrameFlow Auxiliary Losses}. In addition to the flow matching and FAPE losses, we also include the auxiliary losses from \citet{frameflow}. Following \citet{foldflow1}, we apply these atomic constraints for $t < 0.25$, to encourage a better local neighbour representation during the last steps of the generation procedure. These contrainsts consist a regression on the positions of the backbone atoms $\mathcal{L}_{\text{bb}}$ and a pairwise distance loss on the local neighborhood $\mathcal{L}_{\text{2D}}$.

\begin{equation}
    \mathcal{L}_{\text{aux}} = \mathbb{E}_{\mathbb{Q}} \left[ \mathcal{L}_{\text{bb}} + \mathcal{L}_{\text{2D}} \right], \quad
\mathcal{L}_{\text{bb}} = \frac{1}{4N} \sum \| A_0 - \hat{A}_0 \|^2, \quad
\mathcal{L}_{\text{2D}} = \frac{\left\| \mathds{1}_{\{D < 6\hat{A}\}} (D - \hat{D}) \right\|^2}{\sum \mathds{1}_{D < 6\hat{A}} - N}
\end{equation}

where $\mathbb{Q}$ is defined as the factorized joint distribution $\mathbb{Q}(t, x_0, x_1, \tilde{x}_t) := \mathcal{U}(0,1) \otimes \bar{\pi}(x_0, x_1) \otimes \rho_t(\tilde{x}_t | x_0, x_1)$, $\mathds{1}$ is the indicator function that indicates the membership to the local neighborhood according to atomic distances in Angstroms (\AA) and $D$ is a multidimensional array constructed from the pairwise distances between the four heavy atoms belonging to the backbone $D_{ijab} = \| A_{ia} - A_{jb} \|$.

\end{document}